\documentclass[conference]{IEEEtran}
\IEEEoverridecommandlockouts
\usepackage{cite}
\usepackage{amsmath,amssymb,amsfonts}
\usepackage{commath}
\usepackage{algorithmic}
\usepackage[numbered]{bookmark}
\usepackage{graphicx}
\usepackage{textcomp}
\usepackage{ amssymb }
\usepackage{array}
\usepackage{float}
\usepackage{xcolor}
\def\BibTeX{{\rm B\kern-.05em{\sc i\kern-.025em b}\kern-.08em
    T\kern-.1667em\lower.7ex\hbox{E}\kern-.125emX}}

\makeatletter
\newcommand{\thickhline}{%
    \noalign {\ifnum 0=`}\fi \hrule height 1pt
    \futurelet \reserved@a \@xhline
}
\newcolumntype{"}{@{\hskip\tabcolsep\vrule width 1pt\hskip\tabcolsep}}
\makeatother

\begin{document}
\setlength{\parskip}{0.5em}
\title{Anyone GAN Sing}

\author{\IEEEauthorblockN{\textbf{Shreeviknesh Sankaran}\textsuperscript{$1$}, \textbf{Sukavanan Nanjundan}\textsuperscript{$1$}, \textbf{G. Paavai Anand}}
\IEEEauthorblockA{\textit{Department of Computer Science and Engineering,} \\
\textit{SRM Institute of Science and Technology} \\
Chennai, India \\
shreeviknesh@gmail.com, n.sukavanan@gmail.com, paavaiag@srmist.edu.in}
\thanks{\textsuperscript{$1$}co-first authors}
}

\maketitle

\begin{abstract}
The problem of audio synthesis has been increasingly solved using deep neural networks. With the introduction of Generative Adversarial Networks (GAN), another efficient and adjective path has opened up to solve this problem. In this paper, we present a method to synthesize the singing voice of a person using a Convolutional Long Short-term Memory (ConvLSTM) based GAN optimized using the Wasserstein loss function. Our work is inspired by WGANSing by Chandna et al. Our model inputs consecutive frame-wise linguistic and frequency features, along with singer identity and outputs vocoder features. We train the model on a dataset of 48 English songs sung and spoken by 12 non-professional singers. For inference, sequential blocks are concatenated using an overlap-add procedure. We test the model using the Mel-Cepstral Distance metric and a subjective listening test with 18 participants.
\end{abstract}


\section{Introduction} \label{sec:intro}
The problem of singing voice synthesis is similar to that of Text-to-Speech (TTS) synthesis, but the former is much more complicated than the latter. The complexity mainly arises from trying to mimic an extensive range of pitches and phonemes involved in the process of singing. TTS synthesis is primarily controlled by the words or syllables from the text. On the other hand, singing voice synthesis is controlled by a score component in addition to the syllables from the lyrics of the song. The score component determines the pitch and timing of the syllables from the lyrics; in other words, the score defines the flow of a song. 

There are several models (Chandna et al., 2019 \cite{chandna19}, Blaauw et al., 2019 \cite{blaauw17}, Hono et al., 2019 \cite{hono19}, Kaewtip et al., 2019 \cite{kaewtip19}, Lee et al., 2019 \cite{lee19} and Tamaru et al., 2019 \cite{tamaru19}) that have successfully demonstrated the ability to synthesize singing voices of different test subjects. Our model is inspired by WGANSing: A Multi-Voice Singing Voice Synthesizer Based on Wasserstein-GAN by Chandna et al.

Generative Adversarial Networks (GANs) have had immense success in modeling the distribution of highly complex data and have produced state-of-the-art results in image generation \cite{reed16,radford15}. GANs have also been used for TTS synthesis and other such audio generation problems \cite{pascual17,yang17}. But, the number of adaptations of GANs in the audio domain is much fewer when compared to the number of adaptations in the computer vision domain.

The singing voice can be considered as a sequence as there is a sequential flow of notes throughout a song. A song can be constructed only if there is some connection between any two notes throughout the song. Notes thrown around haphazardly without any real flow or connection between the notes can't be considered as ``legitimate'' songs, although some people may find that attractive. This connection between notes can be considered as a sequence, and thus the problem of singing voice synthesis can be approached using sequence prediction techniques such as Long Short-Term Memory (LSTM). In this paper, we propose a Convolutional-LSTM (ConvLSTM) based GAN with an architecture inspired by Chandna et al. to synthesize the singing voice of a person. The choice of using LSTMs stems from the fact that they can model and learn long-range dependencies efficiently \cite{hochreiter97}.

Therefore, we present a block-wise generative model trained using the Wasserstein–-GAN framework for singing voice synthesis. The block-wise nature combined with the convolutional network component enables the model to identify temporal dependencies, just like the inter-pixel dependencies that are identified by GANs in the case of image datasets.

\section{GAN and Wasserstein-GAN}
GAN belongs to the generative frameworks class of deep learning. Since their inception, they have been widely used in the computer vision domain to generate synthetic images and videos that are indistinguishable from real samples \cite{karras19,isola17,zhu17,zhang17}. They consist of two networks (adversaries), a generator and a discriminator which are trained simultaneously. The discriminator is trained to distinguish between synthesized data and real data, whereas the generator is trained to fool the discriminator by synthesizing data that resembles real data. Training of GAN can be formulated as a minimax game \cite{goodfellow14}. The discriminator, on the one hand, tries to maximize its reward, and the generator, on the other hand, tries to minimize the discriminator's reward or, in other words, tries to maximize the discriminator's loss. The loss function for GAN is shown in Eq. \eqref{eq:GAN}.

\begin{equation}
\begin{split}
    \mathcal{L}_{GAN} = \min_{G} \max_{D} \mathbb{E}_{x \sim P_{data}(x)}[log D(x)] \\
    +  \mathbb{E}_{z \sim P_{z}(z)}[log(1-D(G(z))]\label{eq:GAN}
\end{split}
\end{equation}
where $G$ denotes the generator, $D$ denotes the discriminator, $x$ is a sample from the real distribution and $z$ is the input to the generator, which may be noise or some conditional input and is taken from the distribution $P_z$.

As pointed out by Arjovsky et al., while GANs have been efficient in generating images and videos, it has been noted that the above minimax loss function can cause the GAN to get stuck in the early stages of training when the job of the discriminator is easy. More problems, such as vanishing gradient, mode collapse and instability, arise. To overcome such difficulties, Wasserstein-GAN (WGAN) can be used \cite{arjovsky17}. WGAN uses Earth-Mover distance as given in Eq. \eqref{eq:EM} to measure divergence between real and generated distributions. Moreover, WGANs use a critic instead of a discriminator. The critic does not classify inputs as real or fake; instead, it just approximates a distance score between two given distributions (here, the real distribution and the generated distribution).

\begin{equation}
   W(\mathbb{P}_r, \mathbb{P}_g) = \mathop{inf}_{\gamma\in\Pi(\mathbb{P}_r, \mathbb{P}_g)} \mathbb{E}_{(x,y)\sim\gamma}( \norm{x-y} )
   \label{eq:EM}
\end{equation}

The critic can be optimally trained and it does not saturate, thus converging to a linear function. The loss function for WGAN is shown in Eq. \eqref{eq:WGAN}.

\begin{equation}
    \mathcal{L}_{WGAN} = \min_{G} \max_{D} \mathbb{E}_{y \sim P_{r}}[D(y)]
    -  \mathbb{E}_{x \sim P_{x}}[D(G(z))]\label{eq:WGAN}
\end{equation}

The loss functions for both the critic and the generator become deceptively simple. The critic tries to maximize Eq. \eqref{eq:critic} -- i.e., it tries to maximize the difference between its output for real data and its output for synthesized data. The generator tries to maximize Eq. \eqref{eq:gen} -- i.e., it tries to maximize the critic's output for fake or synthesized data.

\begin{equation}
   \mathcal{L}_C = D(x) - D(G(z))\label{eq:critic}
\end{equation}
\begin{equation}
    \mathcal{L}_G = D(G(z))\label{eq:gen}
\end{equation}
where $D(x)$ represents the critic's output for a real instance, $G(z)$ represents the generator's output when given noise $z$ as input or any other conditional input $z$ and $D(G(z))$ represents the critic's output for a fake instance.

We use an extension of GAN model called Conditional GAN (CGAN) which takes an additional conditional vector as input \cite{mirza14}. Adding this conditional vector controls the output and guides the generator in modeling a probability distribution controlled by the vector. The framework is mentioned in Fig. \ref{fig:propsys} in Sec. \ref{sec:propsys}. We use the same training algorithm mentioned in the GAN paper by Goodfellow et al.

\section{LSTM and Convolutional-LSTM}
\begin{figure}[h]
\centering
\includegraphics[width=0.9\columnwidth]{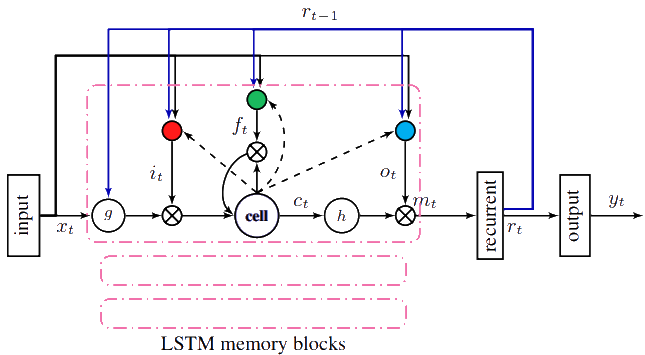}
\caption{A basic LSTM cell.}
\label{fig:lstm}
\end{figure}

Long Short-Term Memory (LSTM) is an Recurrent Neural Network (RNN) architecture that has been extensively used for various applications in Natural Language Processing (NLP) such as speech recognition and semantic parsing \cite{graves05}. LSTMs are capable of learning order dependence and long-range dependencies in sequence prediction problems \cite{hochreiter97}. An LSTM unit is composed of a cell, an input gate, an output gate and a forget gate as shown in Fig. \ref{fig:lstm}. The cell remembers values over arbitrary time intervals and the three gates regulate the flow of information into and out of the cell.

\begin{figure}[h]
\centering
\includegraphics[width=0.9\columnwidth]{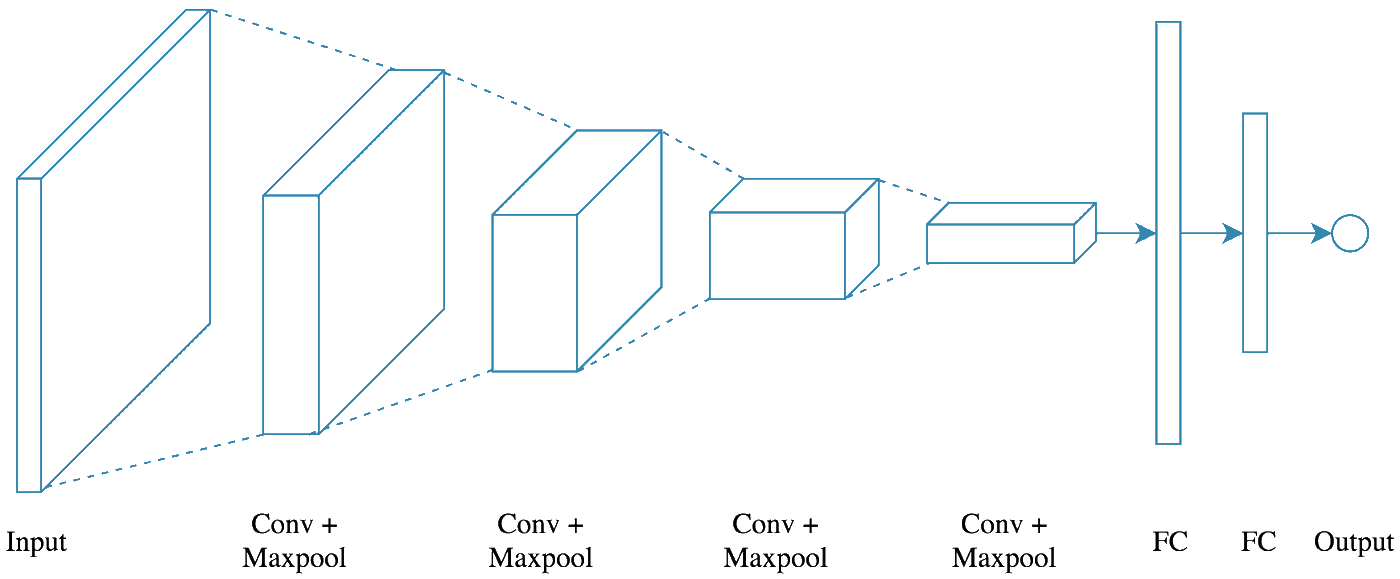}
\caption{A simple CNN architecture.}
\label{fig:cnn}
\end{figure}

On the other hand, a Convolutional Neural Network (CNN) is a deep learning algorithm that is predominantly used in computer vision applications \cite{lecun99}. CNN is a regularized version of multilayer perceptron that is capable of efficiently extracting features and learning them. There are two parts to a CNN: convolution layers and a fully connected neural network that uses the output of the convolutions to predict the output. An example CNN is shown in Fig. \ref{fig:cnn}.

\begin{figure}[h]
\centering
\includegraphics[width=0.76\columnwidth]{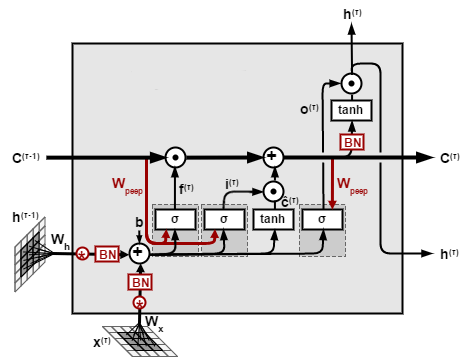}
\caption{A ConvLSTM cell.}
\label{fig:convlstm}
\end{figure}

Convolutional Long Short-Term Memory (ConvLSTM) is an LSTM cell containing a convolution operation embedded inside it as shown in Fig. \ref{fig:convlstm}. It is an LSTM architecture designed explicitly for sequence prediction problems with spatial data, such as images or videos \cite{shi15}.

To take advantage of the abilities of both LSTM and CNN, we use ConvLSTM. ConvLSTM networks are capable of learning long-range dependencies and extracting important features from data, both of which are required in the problem of singing voice synthesis. 

\section{Related Work}
The Neural Parametric Singing Synthesizer (NPSS) by Blaauw et al., is a modified version of WaveNet \cite{oord16} which uses autoregressive architecture. The model features are produced by a parametric vocoder that separates the influence of pitch and timbre. As a result, this helps in training the model with datasets of comparatively smaller size while producing high-quality results, which are comparable to or sometimes even better than state-of-the-art concatenative methods.

Hono et al. present two methods for singing voice synthesis: one is a GAN-based architecture and the other is a conditional GAN-based architecture. This models the inter-frame dependencies as opposed to the inter-block dependencies that are modeled by our model. This difference helps our model to produce more robust results than that of the model presented by Hono et al.

WGANSing by Chandna et al., which is the inspiration for our model, presents a multi-singer singing voice synthesizer. It uses an encoder-decoder based schema for the generator and an encoder schema for the discriminator network. The model produces results that are comparable to that of state-of-the-art models (NPSS in this case). They mention that the synthesis quality can be improved by using previously predicted block of features as a condition to the current batch of features which we have done so by using a ConvLSTM based GAN architecture for singing voice synthesis.

\section{Proposed System} \label{sec:propsys}
We adopt the same architecture used in the WGANSing paper, which is similar to the DCGAN. The main reason for this choice was to establish a baseline and make comparisons easier between models. One main difference between the WGANSing architecture and our architecture is that our generator network uses ConvLSTM layers instead of the CNN layers.

\begin{figure}[h]
\centering
\includegraphics[width=0.9\columnwidth]{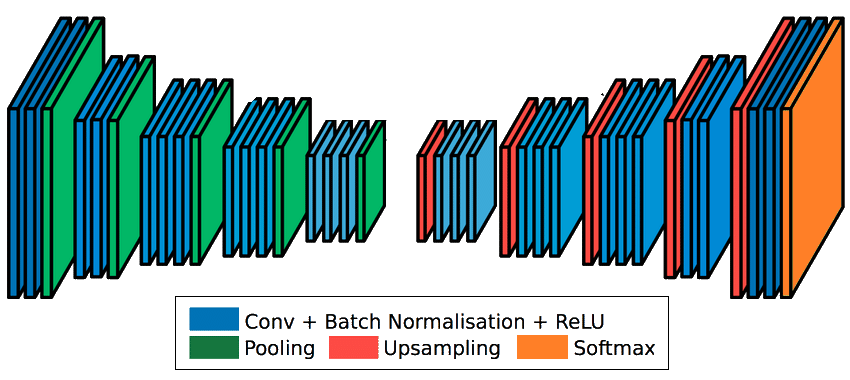}
\caption{The encoder-decoder architecture used in the generator network.}
\label{fig:encdec}
\end{figure}

For the generator network, we use an encoder-decoder architecture consisting of 5 ConvLSTM layers each, as shown in Fig. \ref{fig:encdec}. The whole network is similar to the U-Net architecture that is used for Biomedical Image Segmentation \cite{ronneberger15}. For the discriminator, we use the encoder block of the generator network alone. It asserts the presence of dependencies within a block.

In the encoder block, we use fractionally-strided-convolutions instead of deterministic pooling functions. For example, if a 6x6 pixel image is processed by setting the stride to 3 and the kernel to 3x3, the resulting image is 2x2 in resolution. The inverse of this process begins by determining the spatial resolution and then performing the convolution. While it is not a mathematical inverse, the process is still useful in specific encoding mechanisms. Using this method increases the model's expressiveness ability. 

Furthermore, the encoder-decoder schema leads to conditional dependence between the features of the generator output, within the predicted block of features. This approach implies implicit dependence between the features of a single block but not within the blocks themselves. Therefore, for inference, we use overlap-add of consecutive blocks of output features.

\begin{figure}[h]
\centering
\includegraphics[width=0.9\columnwidth]{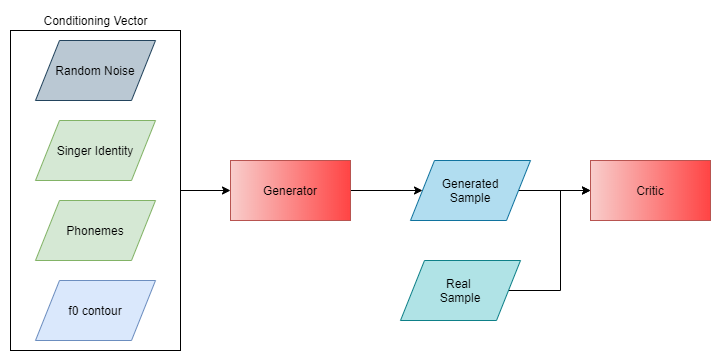}
\caption{The framework for the proposed model. A conditional vector is input to the generator. The critic is trained to identify a real sample.}
\label{fig:propsys}
\end{figure}
As shown in Fig. \ref{fig:propsys}, the generator network inputs a conditional vector consisting of random noise, identity, phonemes as a one-hot vector and f0 contour. Using this conditional GAN, the singing voice of a person is generated.

\section{Dataset}
The dataset used is the NUS-48E Sung and Spoken Lyrics Corpus developed at the Sound and Music Computing Laboratory at the National University of Singapore \cite{duan13}. The corpus is a 169-min collection of recordings of the sung and spoken lyrics of 48 (20 unique) English songs by 12 non-professional singers and a complete set of transcriptions and manual duration annotations at the phone-level for all recordings of sung lyrics, comprising of a total of 25,474 phone instances. 

The corpus consists of 12 folders, one for each subject. Each of these folders consists of ``sing'' and ``read'' folders which consist of 4 sung and corresponding spoken .wav files and their time-aligned phone-level annotations in .txt files. The .wav and .txt files are converted into .hdf5 (hierarchical data format) files to make them easily accessible in Python. These files contain the phonemes and features of each corresponding .wav and .txt files, and thus the features can be used as inputs for the model.

\section{Evaluation Methodology}
For objective evaluation, we use the Mel-Cepstral Distance metric \cite{kubichek93} as shown in Eq. \eqref{eq:mcd} and the results are presented in Tab. \ref{tab:res}. For subjective evaluation, we asked participants to listen to the songs generated by both the models and evaluate them on Audio Quality, Intelligibility and Overall Score. We compared our model to the WGANSing model, both trained on the same dataset and for the same number of epochs -- 750. 

\begin{equation}
    MC = \frac{10 \sqrt{2}}{\ln 10} \frac{1}{T} \sum_{t=1}^T \sqrt{\sum_{i}^{25} \left(C_{ti} - \hat{C}_{ti}\right)^2}\label{eq:mcd}
\end{equation}

We chose a total of 6 songs for the listeners, two songs of each gender without any voice change, two songs of each gender with voice change among the same gender and, two songs of each gender with voice change among opposite genders -- i.e., male voiced song synthesized with a female voice and vice-versa. 

\section{Results}
\begin{figure}[h]
\centering
\includegraphics[width=\columnwidth]{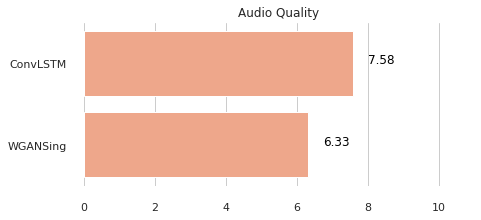}
\caption{Subjective test results for Audio Quality.}
\label{fig:res1}
\end{figure}

\begin{figure}[h]
\centering
\includegraphics[width=\columnwidth]{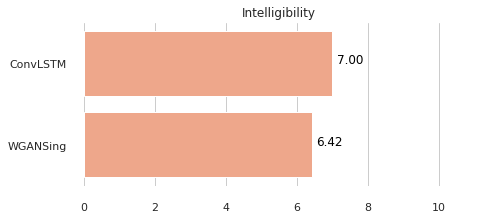}
\caption{Subjective test results for Intelligibility.}
\label{fig:res2}
\end{figure}

\begin{figure}[H]
\centering
\includegraphics[width=\columnwidth]{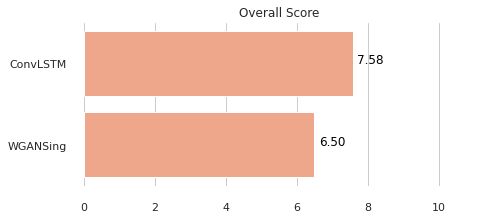}
\caption{Subjective test results for Overall Score.}
\label{fig:res3}
\end{figure}

There were a total of 18 participants, who were all non-native English speakers and with ages in the range 20-22 in our study. The results of the study are shown in Figs. \ref{fig:res1}, \ref{fig:res2} and \ref{fig:res3}.
From the above figures, it is observed that our model performs slightly better than the WGANSing model on all three attributes. This result is further corroborated by the objective measure presented in Tab. \ref{tab:res}.

From Figs. \ref{fig:res1} and \ref{fig:res2}, it is also observed that while both models scored similarly on intelligibility, our model performs better when compared to WGANSing in terms of audio quality. This phenemenon can be mainly attributed to the fact that the songs generated with WGANSing had considerable noise between words or during pauses in the song. However, no such noise was heard in the songs generated by our model. 

It is also observed that the model's performance without voice change was better than the model's performance with voice change. The model's performance further declined when there was a voice change between different genders. Yet, even during the voice change, our model's performance was better than that of the WGANSing model. 

\begin{table}[h]
    \caption{MCD Results}
    \centering
    \begin{tabular}{c||c|c}
        \thickhline
        \raisebox{1em}
        
        Song & ConvLSTM model & Base model \\ [0.5ex]
        \hline
        
        \raisebox{3ex}
        
        MPUR 03 & 18.7567 dB & 21.0440 dB \\ 
        SAMF 13 & 14.3638 dB & 14.6363 dB \\ [0.5ex]
        \thickhline
    \end{tabular}
    
\label{tab:res}
\end{table}

\section{Conclusion}
We have presented a multi-singer singing voice synthesizer using a ConvLSTM-based-conditional-GAN to model a block-wise sequence prediction problem. As this model is inspired by  Chandna et al., we use the same block-wise methodology and architecture as used in their paper for the sake of comparison and testing. While our model seems to perform slightly better than the WGANSing model, both models seem to suffer in certain aspects such as lower audio quality and intelligibility in areas of high notes or pitch, and lower performance when there is a voice change, especially when there is a voice change between opposite genders.

On the whole, using LSTM cells along with convolutions have proven to be an improvement to the WGANSing model, which only had convolutions in an encoder-decoder based architecture. The improvement is mainly because of LSTM's sequence prediction capabilities and as mentioned in Sec. \ref{sec:intro}, song synthesis can be modeled as a sequence prediction problem. 

We believe that the synthesis can be further improved by using algorithms to calculate and model an optimal match between two temporal sequences such as Dynamic Time Warping (DTW). This belief stems from the fact that there will be temporal misalignment between multiple sequences because of acceleration and deceleration during the course of an observation. For instance, using a nearest-neighbor classifier using DTW as the distance measure could improve performance significantly.

\end{document}